\documentclass[10pt,letterpaper]{article}
\usepackage{opex3}
\usepackage{amsmath}
\usepackage{graphicx}

\begin{document}

\title{Polarization modulation time-domain terahertz polarimetry}

\author{C. M. Morris, R. Vald\'{e}s Aguilar, A. V. Stier, N. P.
Armitage}

\address{Department of Physics and Astronomy, The Johns Hopkins University, 3400 N. Charles St., Baltimore, MD 21218}
\begin{abstract}
We present high precision measurements of polarization rotations in the frequency range from 0.1 to 2.5 THz using a polarization
modulation technique. A motorized stage rotates a polarizer
at $\sim80$ Hz, and the resulting modulation of the polarization
is measured by a lock-in technique. We achieve an accuracy of $0.05^{\circ}$
(900 $\mu$rad) and a precision of $0.02^{\circ}$ (350 $\mu$rad) for small
rotation angles. A detailed mathematical description of the technique is presented, showing
its ability to fully characterize  elliptical polarizations from 0.1 to 2.5 THz.
\end{abstract}



Picosecond ($10^{-12}$ s) timescales are one of the most ubiquitous
in condensed matter systems. Scattering times of electrons in metals,
resonant periods of electrons in semiconductors and their nanostructures,
vibrational frequencies of molecular crystals, superconducting {}``Cooper
pair\textquotedblright{} decoherence times, lifetimes of collective
vibrations in biologically important proteins, and even electron transit
times in Intel's THz transistor are all phenomena occurring in the
picosecond range. Such ubiquity makes measurement tools employing
\textit{terahertz} ($10^{12}$ cycles/sec) electromagnetic radiation
potentially quite useful. Unfortunately, measurements in the terahertz
(THz) spectral range have traditionally been challenging to implement
as they lie in the so-called {}``terahertz gap'' - the range of
frequencies above the capabilities of traditional electronics, but
below that of optical generators and detectors (photonics). In recent
years, however, a number of dramatic advances in the form of \textit{time-domain}
THz spectroscopy (TDTS) have helped span this gap, creating an emerging
area of optical and materials research, with a broad range of applications \cite{Tonouchi2007,Nuss1998}.

Photoconductive switch based TDTS works by sequential excitation of
source and detector semiconductor (GaAs in the present work) structures by
a femtosecond pulsed laser. Laser excitation of a biased Auston switch THz
emitter creates an approximately $picosecond$ electromagnetic pulse
(with THz frequency Fourier components) that propagates through
space and interacts with a sample under test. Detection is accomplished
via another photoexcited Auston switch, which is biased across its
electrodes by the transient THz electric field. With these developments,
THz spectroscopy has become a tremendous growth field \cite{DOEreport},
finding potential use in a multitude of areas including materials
characterization for solid-state devices \cite{Kaindl2003,Heyman1998},
optimization of the electromagnetic response of novel coatings \cite{coatings},
probes of superconductor properties \cite{Bilbro2011,Corson1999},
security applications for explosives and biohazard detection \cite{THzBiohazard},
the detection of protein conformational changes \cite{THzproteins},
and non-invasive structural and medical imaging \cite{THzimaging0,THzimaging1,THzimaging3}.

Despite the advances in THz spectroscopy, a number of challenges remain.
For example, highly accurate measurement of polarization states in
the THz range has proven to be difficult \cite{Castro-Camus}. Historically
the highest achievable precisions in the THz or millimeter wave range
have been approximately $1^{\circ}$, measured using rotatable wire grid
polarizers, with higher precisions only being achieved with extremely
specialized setups \cite{Shimano2011}. This is nowhere near the sub-$\mu$rad
resolution that is possible in the near infrared and visible range
\cite{Kapitulnik1994,Xia2006}. A number of groups have developed
polarization sensitive switches \cite{Castro-Camus2005,Makabe2007}
fabricated in a multi-pole configuration, where a single device is
sensitive to two orthogonal electric field polarizations. This simultaneous
detection of both electric field polarizations is advantageous, as
the phase sensitive nature of TDTS means that unlike conventional
polarimetry, only two orthogonal directions have to be measured to resolve
the complete Jones matrix for a polarized wave.

Although multi-pole devices \cite{Castro-Camus2005,Makabe2007}
and calibrated wire-grid polarizer measurements \cite{Neshat12a}
are capable of angular precision of 0.2-0.3$^{\circ}$,
one might hope for even higher angular precision, as many material
systems exhibit very small Kerr or Faraday rotations that are of fundamental
interest for condensed matter physics \cite{Xia2006,Qi2008,Nandkishore2012,Tse11a,Maciejko10a}.
The possibility of even higher precision THz polarimetry would also
open the door for entirely new spectroscopic techniques such as THz
ellipsometry \cite{Shan2009}. Currently the best THz ellipsometers
(based on continuous wave sources) cannot take reliable
data below $\sim3$ THz \cite{Bernhard2004}.

With these ends in mind we have investigated and applied a polarization
modulation technique using a fast rotator in combination with photoconductive
switch based time-domain THz spectroscopy . This technique, originally
developed for use with continuous wave gas lasers and Fourier transform
infrared spectroscopy \cite{Grayson2002,Jenkins2010}, shows a distinct
advantages when applied to TDTS, namely that the phase sensitivity
of TDTS allows the unambiguous measurement of the Jones vector polarization
of a THz wave. The technique has recently been applied to TDTS using
electro-optic detection on topological insulators \cite{RVA2012,George12}.
We find that when used with photoconductive switch TDTS it is capable
of an unprecedentedly high angular precision of better than $0.02^{\circ}$,
with an accuracy of $0.05^{\circ}$ in the frequency range from 0.1
to 1.25 THz, a sensitivity that makes entirely new classes of measurements
possible. In this paper we lay out the mathematical analysis of a polarization modulation measurement and describe initial experimental characterizations of this technique.

\section{Theory\label{sec:Theory}}

In this technique, a rotating polarizer is used to modulate the polarization
state of the terahertz light. In a manner to be described below, it
allows for direct detection of the vertical ($x$) and horizontal
($y$) components of the electric field after passing through a sample using the in-phase and out-of-phase
channels of a lock-in amplifier, respectively. The advantage of this
technique is three-fold. First, measurement time is reduced as a single
measurement determines both electric field polarization states,
compared to the two measurements required with static polarizers.
Second, because the two polarization channels are measured simultaneously,
time dependent common mode noise may be ratioed out, improving the
signal of measurements such as Kerr and Faraday spectroscopies. Third,
as the experiment is only sensitive to the modulation of the polarization state, many detrimental
effects due to the finite extinction ratio of the analyzing polarizers
cancel out, improving the accuracy of the measurement.

We first describe the mathematical analysis in detail. As usual, a time-domain pulse $E_x(t)$ can be written as a superposition of Fourier components

\begin{equation}
\label{eq:fourier_comps}
E_x(t) = \frac{1}{\sqrt{2\pi}}\int_{-\infty}^{\infty} d \omega e^{i \omega t} E_x(\omega)
\end{equation}

Here, $E_x(t)$ is the purely real electric field propagating through the system, and $E_x(\omega)$ describes the amplitude of the complex Fourier components that comprise $E_x(t)$.The mathematical description of the pulse as it propagate through the system is easiest to perform in terms of these Fourier components.   The propagation of the Fourier components as they travel through the
system can be described with the Jones matrix formalism.   In this formalism a matrix
with (in general) complex elements acts on the Fourier component with electric field amplitude
$\mathbf{E^{0}}$ to produce a final polarization state $\mathbf{E^{f}}$:

\begin{equation}
\left(\begin{array}{c}
E_{x}^{f}\left(\omega\right)\\
E_{y}^{f}\left(\omega\right)
\end{array}\right)=
\left(
\begin{array}{cc}
\tilde{M}_{xx}\left(\omega\right) & \tilde{M}_{xy}\left(\omega\right)\\
\tilde{M}_{yx}\left(\omega\right) & \tilde{M}_{yy}\left(\omega\right)
\end{array}
\right)
\left(
\begin{array}{c}
E_{x}^{0}\left(\omega\right)\\
E_{y}^{0}\left(\omega\right)
\end{array}\right)
\end{equation}

The advantage of this formalism is that a complicated series of optical
elements can be represented as a simple set of matrix multiplications
$M$ on the original electric field Fourier component.

The Jones matrix for a polarizer with its transmission axis oriented
at an angle $\Theta$ with respect to the $x$ axis is given by:

\begin{equation}
P_{\Theta}=\begin{pmatrix}
\cos^{2}\left(\Theta\right) & \cos\left(\Theta\right)\sin\left(\Theta\right)\\
\cos\left(\Theta\right)\sin\left(\Theta\right) & \sin^{2}\left(\Theta\right)
\end{pmatrix}
\end{equation}

 Here, a polarizer with $\Theta=0$ polarizes light in the $x$ direction
and is denoted $P_{x}$, and a polarizer with $\Theta=\pi/2$ is called
$P_{y}$. For a rotating polarizer with angular velocity $\Omega$,
the angle $\Theta$ is simply replaced by the factor $\Omega t$ to
obtain the polarizer matrix as a function of time.

In general, the goal of ellipsometry is to determine the values of
the complex elements of the Jones transfer matrix for a sample. The
changes in amplitude and phase these elements produce in the electric
field can then be correlated with physical properties of the sample,
such as the conductivity. The complex transfer matrix $T$ for a sample
is represented as:

\begin{equation}
T\left(\omega\right)=\begin{pmatrix}\tilde{t}_{xx}\left(\omega\right) & \tilde{t}_{xy}\left(\omega\right)\\
\tilde{t}_{yx}\left(\omega\right) & \tilde{t}_{yy}\left(\omega\right)
\end{pmatrix}
\end{equation}

\begin{figure}[H t]

\begin{centering}
\includegraphics[width=0.9\textwidth]{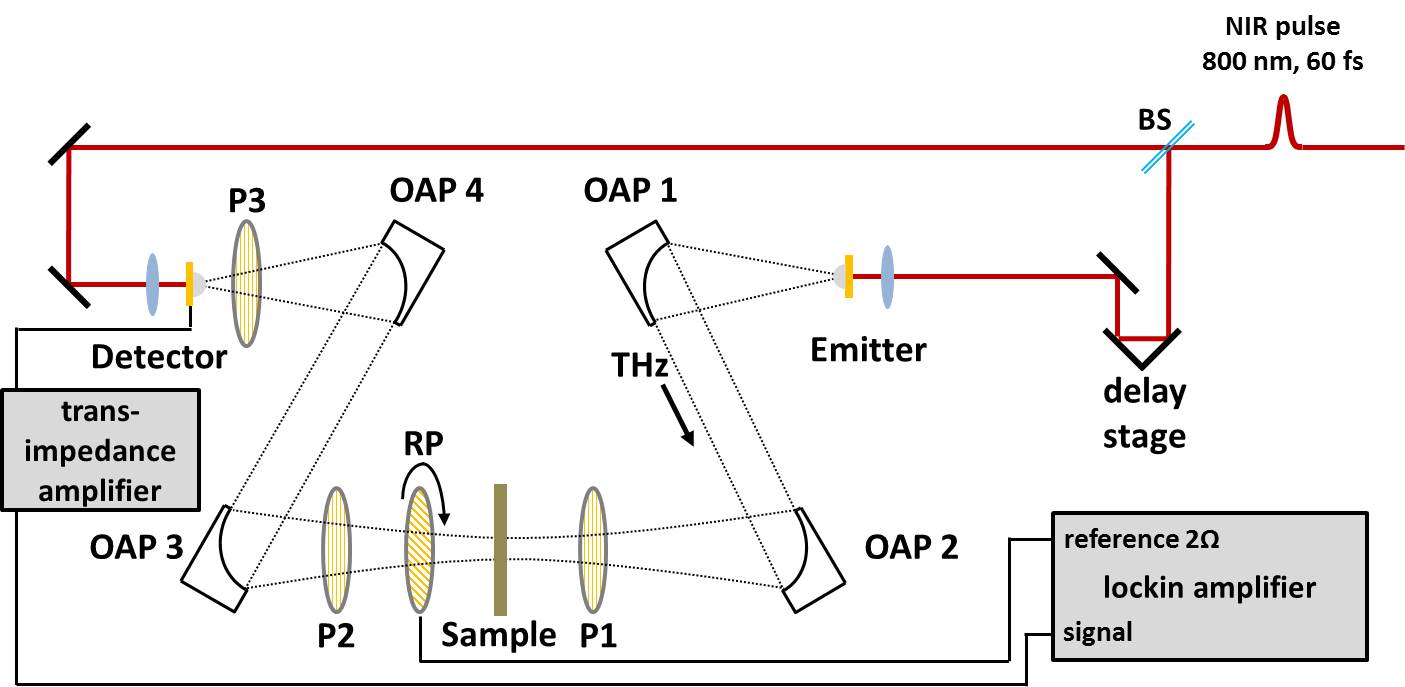} 
\par\end{centering}

\caption{\label{fig:Setup_Diagram}Experimental setup for the fast rotator
measurement. A linearly polarized terahertz waveform passes through
a sample, introducing ellipticity. The rotating polarizer modulates
the polarization, and the modulation is detected by the lock-in amplifier.}
\end{figure}

In the setup described here, a rotating polarizer is used to modulate
the polarization of a terahertz waveform after it has passed through
the sample (Fig. \ref{fig:Setup_Diagram}). Analysis of the time dependence
of the waveform can be used to extract the frequency dependent components
of the $T$ matrix. 

Using the Jones matrices to describe the frequency domain Fourier components of the waveform as they travel through the sample and rotator, the
result is:

\begin{eqnarray}
\begin{pmatrix}E_{x}^{f}\\
E_{y}^{f}
\end{pmatrix} & = & P_{\Omega t}\cdot T\cdot\begin{pmatrix}E_{x}^{0}\\
E_{y}^{0}
\end{pmatrix}\nonumber \\
 & = & \left(\begin{array}{c}
E_{x}^{0}\left[\tilde{t}_{xx}\cos^{2}\left(\Omega t\right)+\tilde{t}_{yx}\cos\left(\Omega t\right)\sin\left(\Omega t\right)\right]\\
E_{x}^{0}\left[\tilde{t}_{xx}\cos\left(\Omega t\right)\sin\left(\Omega t\right)+\tilde{t}_{yx}\sin^{2}\left(\Omega t\right)\right]
\end{array}\right.\nonumber \\
 &  & \left.\begin{array}{c}
+E_{y}^{0}\left[\tilde{t}_{xy}\cos^{2}\left(\Omega t\right)+\tilde{t}_{yy}\cos\left(\Omega t\right)\sin\left(\Omega t\right)\right]\\
+E_{y}^{0}\left[\tilde{t}_{xy}\cos\left(\Omega t\right)\sin\left(\Omega t\right)+\tilde{t}_{yy}\sin^{2}\left(\Omega t\right)\right]
\end{array}\right)
\end{eqnarray}

These electric fields have a mixed dependence on the components of
the $T$ matrix. The analysis can be greatly simplified by adding
vertical polarizers before and after the rotator and sample, respectively, (Fig.
\ref{fig:Setup_Diagram}) to isolate $\tilde{t}_{xx}$ and $\tilde{t}_{yx}$:

\begin{eqnarray}
\begin{pmatrix}E_{x}^{f}\\
E_{y}^{f}
\end{pmatrix} & = & P_{x}\cdot P_{\Omega t}\cdot T\cdot P_{x}\cdot\begin{pmatrix}E_{x}^{0}\\
E_{y}^{0}
\end{pmatrix}\nonumber \\
 & = & \begin{pmatrix}E_{x}^{0}\left(\tilde{t}_{xx}\cos^{2}\left(\Omega t\right)+\tilde{t}_{yx}\cos\left(\Omega t\right)\sin\left(\Omega t\right)\right)\\
0
\end{pmatrix}\nonumber \\
 & = & \begin{pmatrix}\frac{E_{x}^{0}}{2}\left(\tilde{t}_{xx}\left(1+\cos\left(2\Omega t\right)\right)+\tilde{t}_{yx}\sin\left(2\Omega t\right)\right)\\
0
\end{pmatrix}
\label{eq:real_px}
\end{eqnarray}

\noindent or alternatively adding horizontal polarizers to isolate $\tilde{t}_{xy}$
and $\tilde{t}_{yy}$:

\begin{eqnarray}
\begin{pmatrix}E_{x}^{f}\\
E_{y}^{f}
\end{pmatrix} & = & P_{y}\cdot P_{\Omega t}\cdot T\cdot P_{y}\cdot\begin{pmatrix}E_{x}^{0}\\
E_{y}^{0}
\end{pmatrix}\nonumber \\
 & = & \begin{pmatrix}0\\
\frac{E_{y}^{0}}{2}\left(\tilde{t}_{xy}\sin\left(2\Omega t\right)+\tilde{t}_{yy}\left(1-\cos\left(2\Omega t\right)\right)\right)
\end{pmatrix}\label{eq:real_py}
\end{eqnarray}

Note that photoconductive antennas themselves have polarization sensitivity and naturally the best results will be obtained if they are oriented along the direction defined by the static polarizers.

To understand the detection of the waveforms by the photoconductive antennas, there are two relevant times to keep track of in the experiment.  First, there is the time associated with the position of the delay stage (Fig. \ref{fig:Setup_Diagram}), determining the small time window of the terahertz pulse that is being measured. We denote this time as $t_{p}$. This is connected to the position of the delay stage $L$ by $t_p=2L/c$. Second, there is the time associated with the rotation of the polarizer, denoted as $t_{r}$. This is the time that is referred to in Eqs. (\ref{eq:real_px}) and (\ref{eq:real_py}).  We can now rewrite Eq. (\ref{eq:fourier_comps}) for the purely real electric field measured at the detector using this notation:

\begin{equation}\
\label{Fourier_rewrite}
E^f_x(t_p,t_{r}) = \frac{1}{\sqrt{2\pi}} \int_{-\infty}^{\infty} d \omega e^{i \omega t_p} E^f_x(\omega,t_{r})
\end{equation}

\noindent where $E^f_x(\omega,t_{r})$ is defined in Eq. (\ref{eq:real_px}). In the lock-in amplifier this signal is then mixed with the second harmonic of $\Omega$ and integrated for an amount of time determined by the lock-in time constant, $\tau$. The signal of the in-phase lock-in channel is given by

\begin{eqnarray}
S_X\left(t_p\right) & = & \frac{1}{\tau}\int_{0}^{\tau} dt_{r} \cos\left(2\Omega t\right) E^f_x(t_p,t_{r}) \nonumber \\
                              & = &  \frac{1}{\tau \sqrt{2\pi}} \int_{0}^{\tau} dt_{r} \cos\left(2\Omega t\right) \int_{-\infty}^{\infty} d \omega' e^{i \omega' t_p} E^f_x(\omega',t_{r})
\end{eqnarray}

Taking the Fourier transform of $S_X\left(t_p\right)$ allows the Fourier components to be extracted:

\begin{eqnarray}
S_X\left(\omega\right) & = &  \frac{1}{2\pi \tau} \int_{0}^{\tau} dt_{r} \cos\left(2\Omega t\right) \int_{-\infty}^{\infty} dt_p e^{-i \omega t_p} \int_{-\infty}^{\infty} d \omega' e^{i \omega' t_p} E^f_x(\omega',t_{r}) \nonumber \\
                              & = &  \frac{1}{\tau} \int_{0}^{\tau} dt_{r} \cos\left(2\Omega t\right) E^f_x(\omega,t_{r})
\end{eqnarray}

Substituting in for $E^f_x(\omega,t_{r})$ from Eq. (\ref{eq:real_px}):

\begin{eqnarray}
\label{Sx_meas}
S_{X}\left(\omega\right)  & = & \frac{R_{0}\left(\omega\right)}{\tau} \int_{0}^{\tau}dt_{r} \left( \frac{E_{x}^{0}\left(\omega\right) \tilde{t}_{xx}\left(\omega\right)}{2}\left(1+\cos\left(2\Omega t_{r}\right)\right)\right.\nonumber \\
                                         &  & \left.+\frac{E_{x}^{0}\left(\omega\right)\tilde{t}_{yx}\left(\omega\right)}{2}\left(\sin\left(2\Omega t_{r}\right)\right)\right)\cos\left(2\Omega t_{r}\right)\nonumber \\
                                         & = & \frac{R_{0}\left(\omega\right)}{\tau} \left(\frac{E_{x}^{0}\left(\omega\right)\tilde{t}_{xx}\left(\omega\right)}{4}\int_{0}^{\tau} dt_{r} \left(2\cos\left(2\Omega t_{r}\right)+\cos\left(4\Omega t_{r}\right)+1\right)\right.\nonumber \\
                                         &  & \left.+\frac{E_{x}^{0}\left(\omega\right)\tilde{t}_{yx}\left(\omega\right)}{4}\int_{0}^{\tau} dt_{r} \sin\left(4\Omega t_{r}\right)\right)\nonumber \\
                                         & = & \frac{1}{4} R_{0}\left(\omega\right) E_{x}^{0}\left(\omega\right) \tilde{t}_{xx}\left(\omega\right)
\end{eqnarray}

\noindent where $R_{0}\left(\omega\right)$ parameterizes the combined responsivity of the antenna
and lock-in amplifier. In the $y$ channel of the lock-in the signal
is mixed with $\sin\left(2\Omega t_{r}\right)$, and similarly gives:

\begin{equation}
\label{Sy_meas}
S_{Y}\left(\omega\right)=\frac{1}{4} R_{0}\left(\omega\right) E_{x}^{0}\left(\omega\right) \tilde{t}_{yx}\left(\omega\right)
\end{equation}

How do these measured values relate to the quantities we wish to measure,
namely the electric fields after the sample? An analysis of the effect
of the sample on vertically polarized light without the rotator shows
that the electric fields after the sample are:

\begin{equation}
\label{Static_meas}
\begin{pmatrix}E_{x}^{f}\\
E_{y}^{f}
\end{pmatrix}=T\cdot P_{x}\cdot\begin{pmatrix}E_{x}^{0}\\
E_{y}^{0}
\end{pmatrix}=\begin{pmatrix}\tilde{t}_{xx}E_{x}^{0}\\
\tilde{t}_{yx}E_{x}^{0}
\end{pmatrix}
\end{equation}

Comparing the results of Eqs. (\ref{Sx_meas}) and (\ref{Sy_meas}) with the fields found in Eq. (\ref{Static_meas}), it is apparent that the rotator measurement is equivalent to direct measurement of the x and y components of the electric field after the sample simultaneously in the two lock-in channels. Eqs. (\ref{Sx_meas}) and (\ref{Sy_meas}) represent the principle result of this analysis: that a single measurement with the rotator technique is exactly equivalent to two measurements with a standard static polarizer setup.

In a real system, the non-ideality of the components must be taken into account. For example, perfectly aligned off-axis
parabolic mirrors produce no rotation of the polarization state, but
in a real system perfect alignment is impossible to achieve. In this
regard placement of optical elements in the setup is crucial. In Fig.
\ref{fig:Setup_Diagram} off-axis parabolic mirrors (OAPs) 1 and 2
can introduce a rotation $\phi_{12}$ of the initial polarization state of the
electric field from the emitter. To eliminate measurement errors introduced
by this rotation, it is critical to place the first static polarizer
\textit{after} OAPs 1 and 2, which essentially redefines the initial
vertical electric field for the measurement at this point.

For OAPs 3 and 4 the same problem exists. Consider the situation explicitly
if polarizer P2 is placed not immediately before OAP 3, but instead
after OAP 4. The total rotation $\phi_{34}$ produced from small misalignments
in OAPs 3 and 4 can be represented by the rotation matrix:

\begin{equation}
\left(\begin{array}{cc}
\cos\left(\phi_{34}\right) & \sin\left(\phi_{34}\right)\\
-\sin\left(\phi_{34}\right) & \cos\left(\phi_{34}\right)
\end{array}\right)\label{eq:Rotation matrix}
\end{equation}

The $x$ polarized light detected in the two channels of the lock-in
becomes a mixture of the components of the $T$ matrix:

\begin{eqnarray}
S_{X} & = & \frac{R_{0}E_{x}^{0}}{4}\left(\tilde{t}_{xx}\cos\left(\phi_{34}\right)-\tilde{t}_{yx}\sin\left(\phi_{34}\right)\right)\nonumber \\
S_{Y} & = &\frac{R_{0}E_{x}^{0}}{4}\left(\tilde{t}_{yx}\cos\left(\phi_{34}\right)+\tilde{t}_{xx}\sin\left(\phi_{34}\right)\right)
\end{eqnarray}

As before, placing a polarizer between
the rotator and OAP 3 simplifies the result:

\begin{equation}
S_{X}=\frac{R_{0}E_{x}^{0}}{4}\tilde{t}_{xx}\cos\left(\phi_{34}\right),\; S_{Y}=\frac{R_{0}E_{x}^{0}}{4} \tilde{t}_{yx}\cos\left(\phi_{34}\right)
\end{equation}

Now each channel acquires the same rotation factor of $\cos\left(\phi_{34}\right)$,
which merely acts to scale the overall magnitude of the electric field.
There is no mixing of the polarization states in the lock-in channels.

The finite extinction ratio of the polarizers in the system is also
a factor in the overall system performance. The Jones matrix for
an imperfect polarizer is \cite{Huang2010}:

\begin{equation}
\label{finite_polarizer}
P=\begin{pmatrix}\cos^{2}\left(\theta\right)+\eta\sin^{2}\left(\theta\right) & (1-\eta)\cos\left(\theta\right)\sin\left(\theta\right)\\
(1-\eta)\cos\left(\theta\right)\sin\left(\theta\right) & \eta\cos^{2}\left(\theta\right)+\sin^{2}\left(\theta\right)
\end{pmatrix}
\end{equation}

\noindent where $\eta$ is the extinction ratio of the polarizer, defined
as the ratio of the transmitted electric field when the polarizer
transmission axis is perpendicular and parallel to the electric
field direction, $\eta=E_{\bot}^{trans}/E_{\Vert}^{trans}$. Accounting
for the finite extinction ratio ($\eta_r$) of the rotating polarizer in the Jones
matrix analysis, the measured signals become:

\begin{equation}
S_{X}=\frac{R_{0}E_{x}^{0}}{4}\tilde{t}_{xx}\cos\left(\phi_{34}\right) \left(1-\eta_{r}\right),\; S_{Y}=\frac{R_{0}E_{x}^{0}}{4} \tilde{t}_{yx}\cos\left(\phi_{34}\right) \left(1-\eta_{r}\right)
\end{equation}

The finite extinction ratio of the rotating polarizer acts only as
a simple scaling of the amplitude of the measured electric field.
This highlights a distinct advantage of the modulation technique over
the standard static polarizer measurements: only modulated signals
are measured by the lock-in, and non-ideality of the rotating polarizer
only produces a small effect on the amplitude of the measured electric
fields. Commercially available polarizers in this frequency range
typically have extinction ratios of $\eta\sim1/50$, resulting
in only $2\%$ difference between the ideal and real signals. The
best figure of merit for the rotating polarizer becomes its stability
under rotation and its robustness to high rotation rates to reduce
$1/f$ noise, not the extinction ratio.

Finally, the effects of the finite extinction ratios of the static
polarizers must be considered. This changes the form of the vertical
polarization matrix, $P_{x}$, to

\begin{equation}
\left(\begin{array}{cc}
1 & 0\\
0 & 0
\end{array}\right)\rightarrow\left(\begin{array}{cc}
1 & 0\\
0 & \eta
\end{array}\right)\label{eq:finite_extinction}
\end{equation}

Finite extinction ratios of the static polarizers mix different $T$ matrix components in the measured signals. Since the rotator projects the electric field onto the second polarizer at large angles,
a non-trivial electric field leakage results and the elements of the
$T$ matrix are mixed in the two detection channels. The obvious solution
is to increase the quality of polarizer, but while polarizers with
significantly higher extinction ratios in the terahertz range have
been demonstrated, they are still not at the stage of commercial production
\cite{Ren0}. As shown below the solution to this problem relies on
the final analysis performed on the collected time domain waveforms.

As described above, in order to obtain the electric field components as a function of frequency, the
 measured time domain waveform is
Fourier transformed to obtain the frequency response. For many measurements
performed in this system, the quantity of interest is the rotation angle
introduced by the sample, which can be found by simply taking the
ratio of the Fourier transforms of $S_{y}$ and $S_{x}$. For the
case of vertical P1 and P2 polarizers (Eq. (\ref{Sx_meas})): 

\begin{equation}
\frac{S_{y}\left(\omega\right)}{S_{x}\left(\omega\right)}=\frac{\mathrm{\mathit{E_{x}^{0}\left(\omega\right)}\mathit{\tilde{t}_{yx}}\mathit{\left(\omega\right)}}}{\mathrm{\mathit{E_{x}^{0}\left(\omega\right)}\mathit{\tilde{t}_{xx}}\mathit{\left(\omega\right)}}}=\frac{\tilde{t}_{yx}\mathit{\left(\omega\right)}}{\tilde{t}_{xx}\mathit{\left(\omega\right)}}\label{eq:arctangent}
\end{equation}

If the polarizers are assumed to be ideal, simple cases such as rotation
can be explicitly solved. The T matrix for a rotation $\phi$ is given
by Eq. (\ref{eq:Rotation matrix}), which allows Eq. (\ref{eq:arctangent})
to take a simple form

\begin{equation}
\frac{\tilde{t}_{yx}\mathit{\left(\omega\right)}}{\tilde{t}_{xx}\mathit{\left(\omega\right)}}=\frac{\mathrm{\sin\left[\phi\mathit{\left(\omega\right)}\right]}}{\mathrm{\cos\left[\phi\mathit{\left(\omega\right)}\right]}}
\end{equation}

\noindent which easily gives the rotation angle $\phi$

\begin{equation}
\arctan\left(\frac{\mathrm{\sin\left[\phi\mathit{\left(\omega\right)}\right]}}{\mathrm{\cos\left[\phi\mathit{\left(\omega\right)}\right]}}\right)=\phi\left(\omega\right)
\label{eq:simplearctan}
\end{equation}

However a finite extinction ratio of the polarizers complicates Eq. (\ref{eq:arctangent}).
If the finite extinction ratio of the static polarizers in the $P_{x}$ geometry is taken into account,
it is impossible to solve Eq. (\ref{eq:simplearctan}) analytically
for the rotation. The situation is particularly complicated for the case where small misalignments in off-axis parabolic mirrors given finite angles $\phi_{12}$ and $\phi_{34}$.  However, we can perform a Taylor expansion in $\eta$
up to second order for a simple rotation, giving

\begin{eqnarray}
\arctan\left(\frac{S_{y}\left(\omega\right)}{S_{x}\left(\omega\right)}\right) & = & \phi\left(\omega\right) - \left(-\frac{E_{x}^{0}}{E_{y}^{0}+E_{x}^{0}\cot\left(\phi_{12}\right)}\right.\nonumber \\
 &  & \left.+\frac{E_{y}^{0}}{E_{x}^{0}+E_{y}^{0}\tan\left(\phi_{12}\right)}+\tan\left(\phi_{34}\right)\right)\eta\label{eq:expansion}
\end{eqnarray}

This expansion shows that a single measurement is in general not sufficient
to determine the rotation angle. Even with the setup optimized by
proper polarizer placement to eliminate first order rotation effects
due to the off-axis parabolic mirrors ($\phi_{12}$ and $\phi_{34}$)
(as discussed above), the finite extinction ratio of the polarizers
can still cause systematic error. Note however, that the second term
in Eq. (\ref{eq:expansion}) does not depend on the rotation angle $\phi$.
In this regard, two methods can be used to eliminate the systematic
error. If the sample itself can be used to give a reference angle
of rotation (say at a particular temperature or magnetic field), this
reference can be subtracted to eliminate the constant term in Eq. (\ref{eq:expansion})
and obtain the relative \textit{change }in rotation angle. An example
could be comparing a rotation measurement above and below $T_{c}$
in a superconductor for evidence of time reversal symmetry breaking
\cite{Xia2006} or of a magnetic material above and below a magnetic
ordering temperature. Alternatively, if an accurate reference sample
can be used, the sample and reference angle scans can be subtracted,
and systematic errors of this kind can be eliminated. Also note that
in general the contribution of $E_{y}^{0}$ will
be small, as in the typical experimental geometry the vertically oriented photoconductive
switches give an electric field which is primarily oriented along
the $x$ direction.

Up to this point, a technique has been developed and shown to be capable
of measuring simple rotations. What are the ultimate capabilities
of this technique, and how do they connect to full terahertz ellipsometry
measurements? To answer this question, we must first define what is measured in a standard ellipsometric measurement. In standard ellipsometry, light with a known polarization is focused onto a sample, and the reflected or transmitted light is collected by polarization sensitive detectors. The amplitude of the light in the $x$ and $y$ axes is determined as a function of the frequency from this measurement. 

\begin{equation}
\label{eq:rho}
\rho= \frac{E_y\left(\omega\right)}{E_x\left(\omega\right)} = \tan \left[ \psi \left(\omega\right) \right] e^{i \Delta\left(\omega\right)}
\end{equation}

\noindent where $\tan \psi = |E_{y}/E_{x}|$ and $\Delta$ represents the temporal phase difference between the two field components \cite{fujiwara}. The broadband polarization modulation measurement presented here gives $E_x\left(\omega\right)$ and $E_y\left(\omega\right)$, so both $\psi \left(\omega\right)$ and $\Delta \left(\omega\right)$ can be calculated and a full frequency dependent ellipsometry measurement is possible.

Next we can ask, how do these measured complex field components connect to rotations and the ellipticity? In general, fully polarized states of monochromatic transverse electromagnetic
waves trace out an ellipse in time with
major ($a$) and minor ($b$) axes, each at an angle $\theta$ with
the $x$ and $y$ axes in the laboratory frame, respectively \cite{fujiwara}. The
ellipticity is defined by the major and minor axes as $\tan(\epsilon)=\frac{b}{a}$.
For example, for circularly polarized light $a=b$ and $\epsilon=\pm45^{\circ}$
with + for right circular polarization, and - for left. Linearly polarized
light is characterized by $b=0$ and $\epsilon=0$. All other polarization
states are called \emph{elliptical}. For the simple polarizer in Eq. (\ref{eq:simplearctan}), the ellipticity $\epsilon=0$, and the angle $\theta$ corresponds exactly to the angle of rotation $\phi$.

To perform a complete analysis, it is important to note \cite{Born1997} that the ellipse the electric field traces in time is related to the measured ellipsometric quantities $\tan \psi$ and $\Delta$ by

\begin{eqnarray}
\tan(2\theta) & = & \frac{2\tan \psi\cos{\Delta}}{1-\tan^{2} \psi}\label{Born1}\\
\sin(2\epsilon) & = & \frac{2\tan \psi\sin{\Delta}}{1+\tan^{2} \psi}\label{Born2}
\end{eqnarray}

To start the analysis, we take $\arctan \rho$ defined in Eq. (\ref{eq:rho}), which can then be expanded

\begin{eqnarray}
\label{atan_rho}
\arctan \rho & = & \frac{1}{2}\arg\left(1+\tan \psi e^{i\Delta}\right)-\frac{1}{2}\arg\left(1-\tan \psi e^{i\Delta}\right)\nonumber \\
 &  & +i\frac{1}{4}\log{\left(\frac{\tan^{2} \psi \cos^{2}{\Delta}+(1+\tan \psi \sin{\Delta})^{2}}{\tan^{2} \psi \cos^{2}{\Delta}+(1-\tan \psi \sin{\Delta})^{2}}\right)}\label{Expangle}
\end{eqnarray}

First taking the real part of Eq. (\ref{atan_rho}) we find

\begin{eqnarray}
\mathrm{Re}{[\arctan \rho]} & = & \frac{1}{2}\arg{\left(1+\tan \psi e^{i\Delta}\right)}-\frac{1}{2}\arg{\left(1-\tan \psi e^{i\Delta}\right)}\nonumber \\
 & = & \frac{1}{2}\left[\arctan{\left(\frac{\tan \psi \cos{\Delta}}{1-\tan \psi \sin{\Delta}}\right)}+\arctan{\left(\frac{\tan \psi \cos{\Delta}}{1+\tan \psi \sin{\Delta}}\right)}\right]\nonumber \\
 & = & \frac{1}{2}\arctan{\left(\frac{2\tan \psi \cos{\Delta}}{1-\tan^{2} \psi}\right)}\label{eq:Reangle}
\end{eqnarray}

Inserting Eq. (\ref{Born1}) into Eq. (\ref{eq:Reangle}) gives

\begin{eqnarray}
\mathrm{Re}[\arctan \rho]=\mathrm{Re}\left[\arctan\left(\frac{E_y\left(\omega\right)}{E_x\left(\omega\right)}\right)\right]=\frac{1}{2}\arctan{\left(\tan2\theta\right)}=\theta\label{Reangle2}
\end{eqnarray}

\noindent which is simply the rotation angle $\phi$ that was found for the simple polarizer case. Now we find the imaginary part
of Eq. (\ref{Expangle}).

\begin{eqnarray}
\mathrm{Im}[\arctan \rho] & = & \frac{1}{4}\log{\left(\frac{\tan^{2} \psi \cos{\Delta}^{2}+(1+\tan \psi \sin{\Delta})^{2}}{\tan^{2} \psi\cos{\Delta}^{2}+(1-\tan \psi \sin{\Delta})^{2}}\right)}\nonumber \\
 & = & \frac{1}{4}\log{\left[\frac{1+\frac{2\tan \psi\sin{\Delta}}{1+\tan^{2} \psi}}{1-\frac{2\tan \psi\sin{\Delta}}{1+\tan^{2} \psi}}\right]}\nonumber \\
 & = & \frac{1}{2}\mathrm{arctanh}{\left(\frac{2\tan \psi\sin{\Delta}}{1+\tan^{2} \psi}\right)}\label{Imangle}
\end{eqnarray}

Using Eq. (\ref{Born2}) we can simplify:

\begin{eqnarray}
\mathrm{Im}[\arctan \rho]=\frac{1}{2}\mathrm{arctanh}{(\sin{(2\epsilon)})}\label{Imangle2}
\end{eqnarray}

The $\mathrm{Im}\left(\arctan \rho\right)$ can be used to determine the the ellipticity $\epsilon$. This analysis shows that the complex angle extracted from $\arctan \rho$ contains all the information necessary to determine the frequency dependent elliptical polarization state of the final terahertz electric field. The real part is a direct measurement of the rotation angle of the light, and the imaginary part determines the ellipticity of the final electric field. In fact, this analysis does not rely on the polarization modulation technique, but is valid for the static measurement described by Eq. (\ref{Static_meas}) as well.  The advantage of the polarization modulation technique is that the full characterization of the frequency dependent electric field ellipse can be extracted from a single measurement. This shows the power of measuring the time dependent electric field and collecting the full amplitude and phase information of the waveform.

\section{Experiment\label{sec:Experiment}}

\begin{figure}[h b]
 
\begin{centering}
\includegraphics[width=0.48\textwidth]{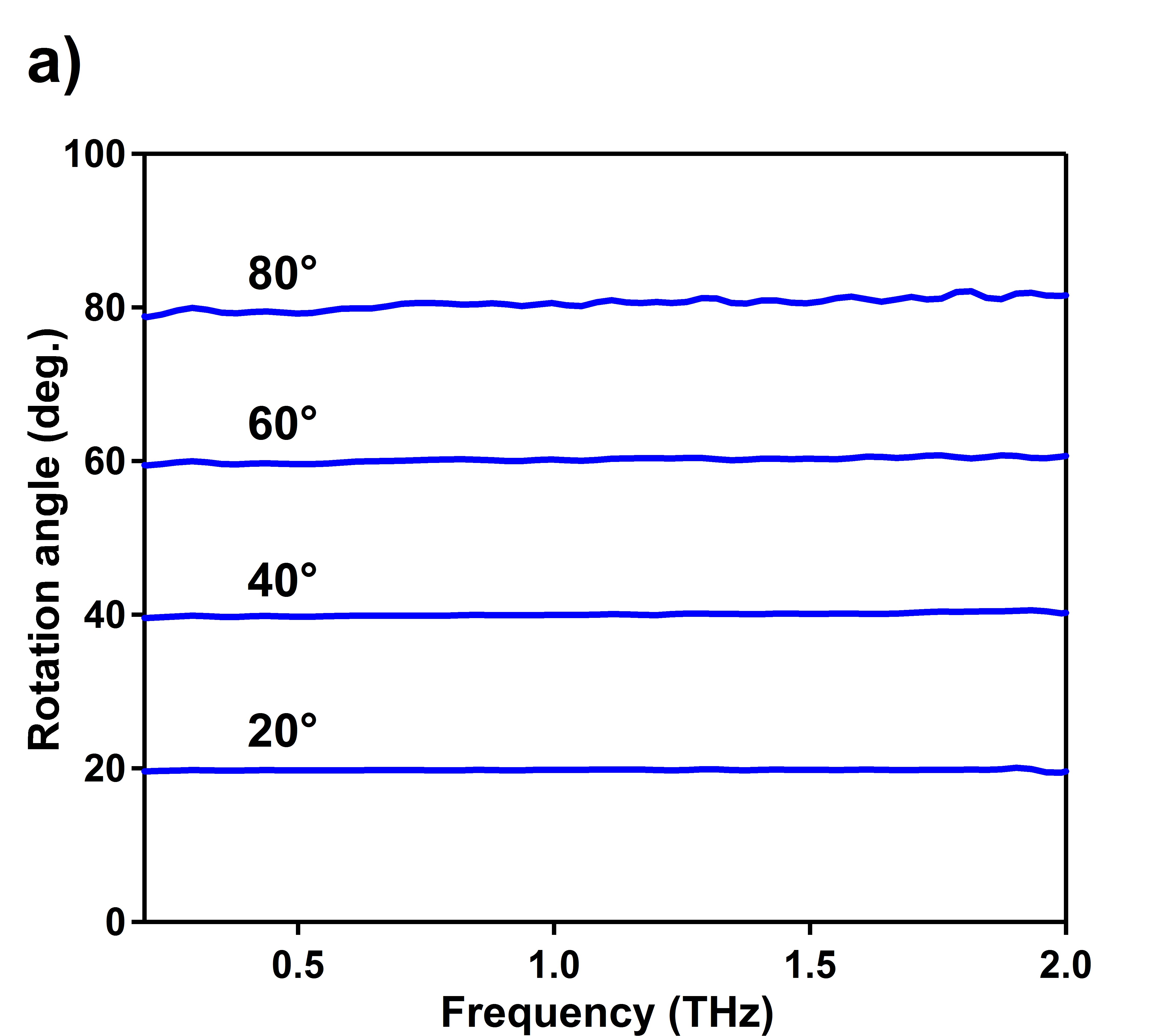}\includegraphics[width=0.48\textwidth]{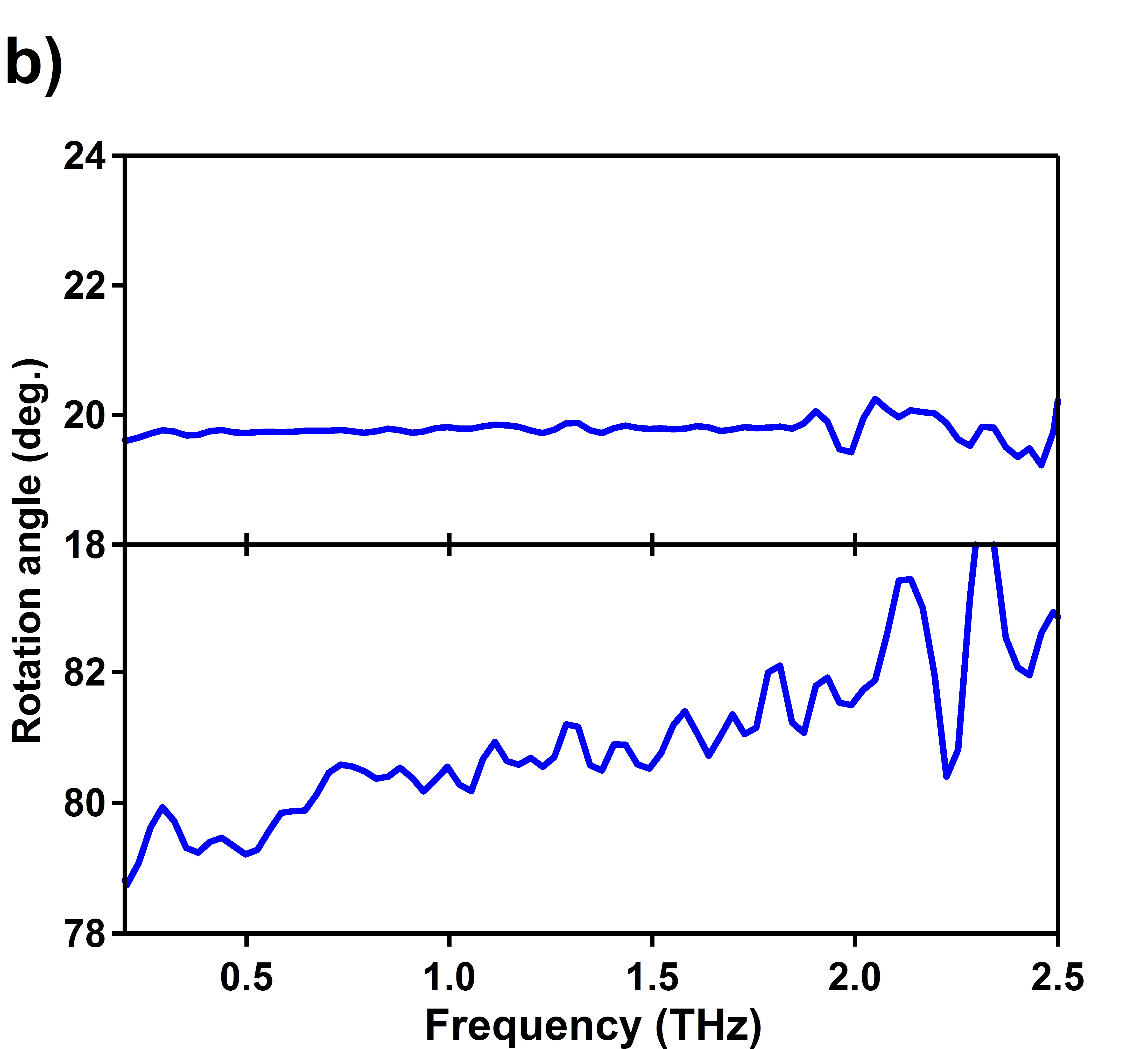} 
\par\end{centering}

\begin{centering}
\includegraphics[width=0.48\textwidth]{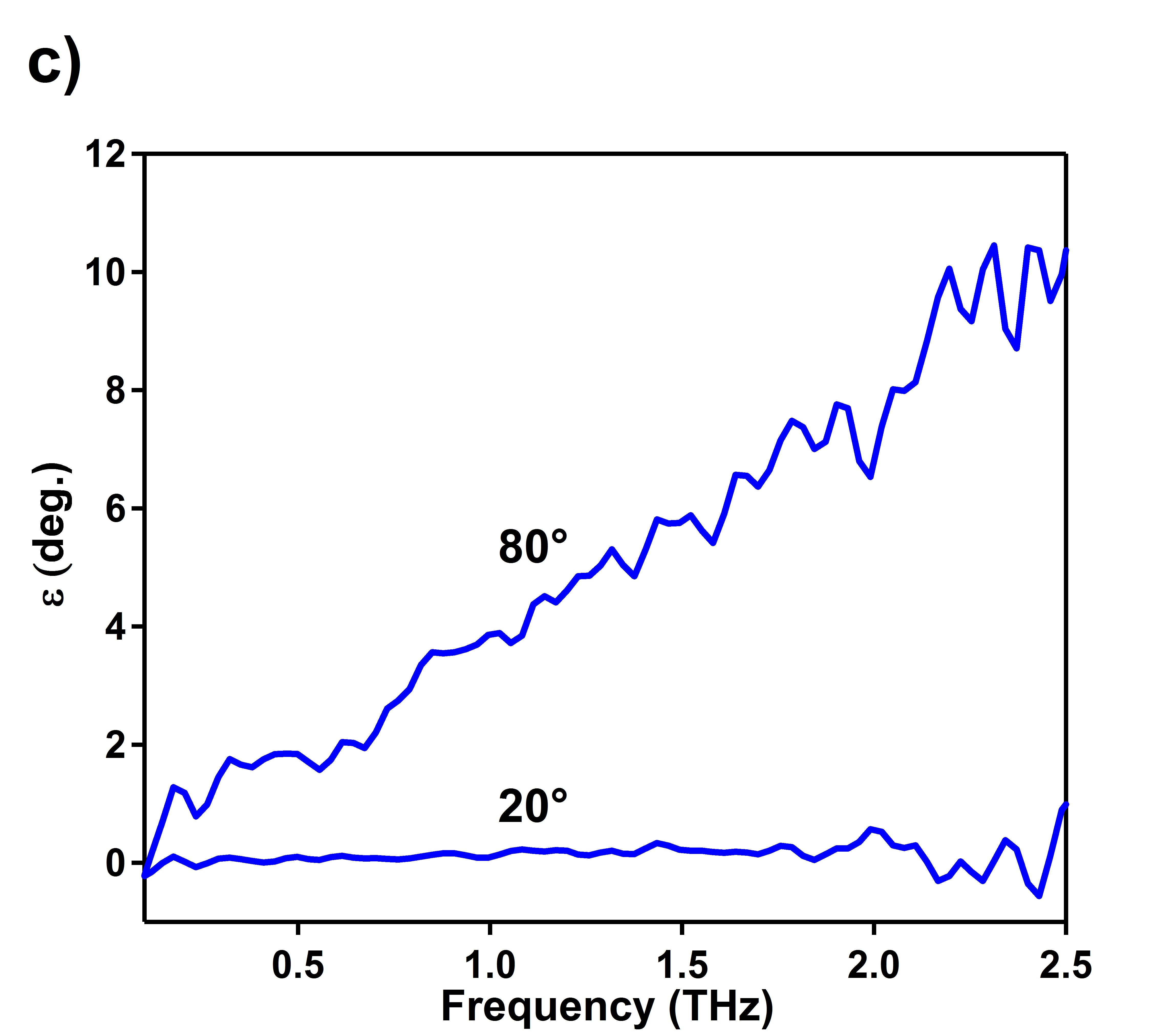}\includegraphics[width=0.48\textwidth]{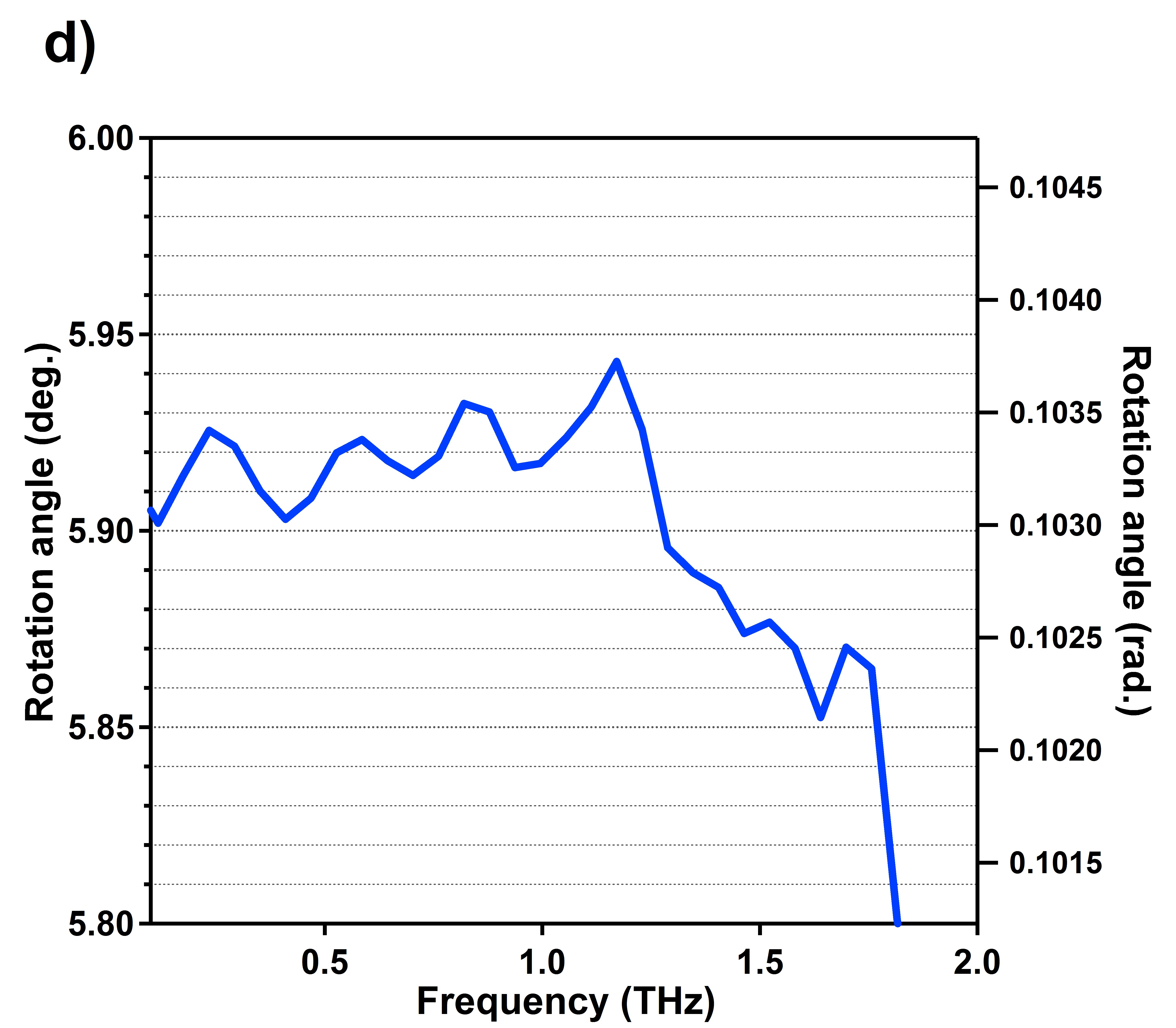} 
\par\end{centering}

\caption{\label{fig:Polarization-angle-characterizat}System resolution characterization.
a) Measurement of test polarizer angles from $20^{\circ}$ to $80^{\circ}$. b) Comparison of the accuracy and
precision for small and large test polarizer angles. c) Ellipticity
angle $\epsilon$ for two polarizer angles. d) A test polarizer angle of $6^{\circ}$, showing $0.05^{\circ}$ accuracy and $0.02^{\circ}$ precision up to 1.25 THz.}
\end{figure}

The experimental setup is shown in Fig. \ref{fig:Setup_Diagram}.
The generation and detection of the terahertz waveform is accomplished
by a standard time-domain terahertz spectrometer using photoconductive
antennas. A $\sim60$ fs pulse from a KM Labs Ti:Sapph oscillator
with an 80 MHz repetition rate is split into two paths via a beamsplitter. The first
pulse strikes a biased photoconductive antenna, generating a terahertz
waveform with a spectral bandwidth from 0.1 to 3 THz. The  OAPs are laid out in an $8f$ geometry. OAPs 1 and 2
are used to collimate and focus the waveform.  A wire grid polarizer
(P1) after OAP 2 defines the initial vertical polarization state of
the system as the terahertz light is focused onto the sample. The
vertically polarized light passes through the sample, where it becomes
elliptically polarized according to the sample transfer ($T$) matrix.
The elliptically polarized terahertz light passes through a rotating
wire-grid polarizer (RP), made by QMC Instruments, which modulates the elliptical polarization
at a frequency $\Omega$. A second vertical polarizer (P2) projects
the rotated light back to the vertical ($x$) axis, and a pair of
off-axis parabolic mirrors recollimate and focus the terahertz light
onto the detector antenna. The second pulse created by the beamsplitter is
used for detection of the waveform. An additional vertical polarizer
(P3) is used to account for any rotations produced by OAPs 3 and 4,
as the antenna has a small, but finite response for the horizontal
polarization, which would result in mixing of the elements of the
$T$ matrix in the detection as shown above. The terahertz electric field
at the antenna produces a small AC current, and a transimpedance amplifier
is used to convert this to a voltage. The voltage signal is detected
by an SRS 830 lock-in amplifier, where it is mixed with the second
harmonic of the rotation frequency $\Omega$. The rotation frequencies
used here (2$\Omega$ between 50 Hz and 80 Hz) are chosen to be high enough to
significantly reduce $1/f$ noise, but small enough to minimize the
possibility of damaging the polarizer.

The rotator was built by the Instrument Development Group in the Johns Hopkins Department of Physics and Astronomy.  The polarizer is held by a 2" aluminum cylinder that can be rotated at high speeds using Bearing Works silicon nitride ball bearings. This is connected via a belt drive to a high speed AC motor, the Faulhaber Minimotor 4490, which has a variable speed controller that can be used to rotate the polarizer up to 2$\Omega$ = 200 Hz (6000 rpm). To generate a trigger signal for the lock-in, two small holes are drilled in the rotating cylinder, and an LED and photodiode are placed on opposite sides of the cylinder. When a hole passes the LED, it triggers the photodiode on the opposite side of the cylinder.  This produces pulses at twice the rotation frequency, giving the 2$\Omega$ signal needed for the lock-in detection.  All this is held in a metal casing that can be securely attached to the optical table to reduce detrimental effects caused by motor vibration.

To characterize system performance, a wire-grid polarizer in a static
rotation mount was used as a reference sample. To demonstrate the
versatility of the technique, a number of polarizer angles were measured
with the results shown in Fig. \ref{fig:Polarization-angle-characterizat}.
For large rotation angles, the precision and accuracy of the measurement
is reduced, as the comparison of the measurements for rotation angles
of $20^{\circ}$ and $80^{\circ}$ (Fig. \ref{fig:Polarization-angle-characterizat}b)
shows. This seeming loss of precision and accuracy has its origins
in the non-ideality of the polarizer as a test sample. The polarizer
does not simply rotate the polarization, but rather projects the polarization
with an angle dependent ($\cos^{2}\theta$) attenuation. For large
angles, the sample polarizer significantly reduces the amplitude of
the initial terahertz electric field, thereby decreasing the achievable
signal to noise. The finite extinction ratio of the polarizers also
becomes more important for larger polarizer angles, as the amplitude
of the orthogonally polarized field transmitting through the polarizer
grows. Additionally, at these large angles the amplitude of the orthogonal
transmitted electric field becomes highly frequency dependent. These
two effects are shown in detail in Fig. \ref{fig:Polarization-angle-characterizat}b.
At $20^{\circ}$, the measured angle is constant to within $0.1{}^{\circ}$
over the range from 0.1-1.75 THz, with a precision of $\sim0.05{}^{\circ}$.
At $80{}^{\circ}$, the angle is only constant to $2{}^{\circ}$ within
the same frequency range, and the precision of the measurement is
reduced to $\sim0.5{}^{\circ}$. The ideal test sample would be a
broadband half waveplate that could rotate without attenuation. However,
broadband terahertz waveplates are at the developmental stage \cite{Masson2006},
so a wire grid polarizer serves as the simplest available test sample.

\begin{figure}[H t]
 
\begin{centering}
\includegraphics[width=0.73\textwidth]{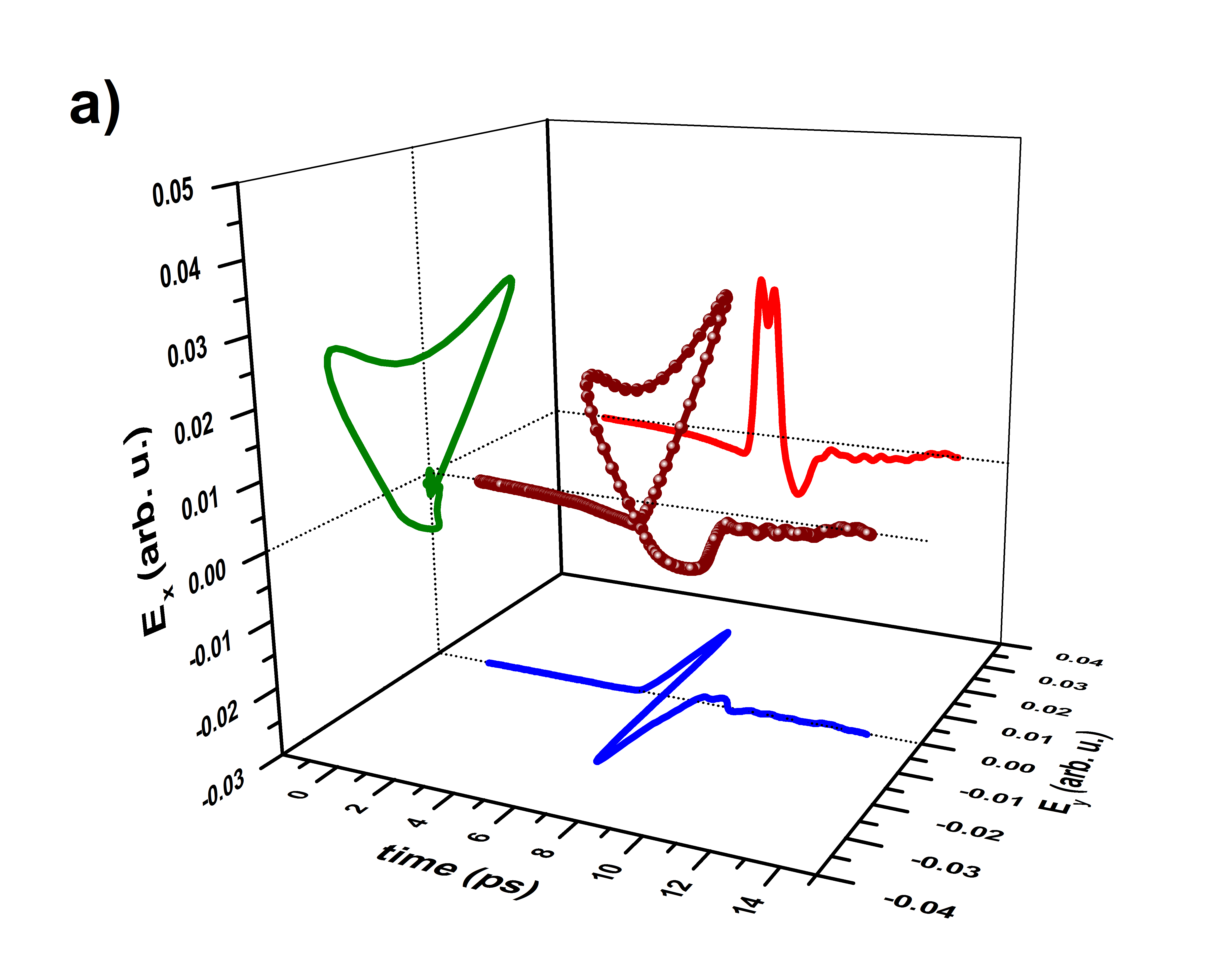} 
\par\end{centering}

\begin{centering}
\includegraphics[width=0.59\textwidth]{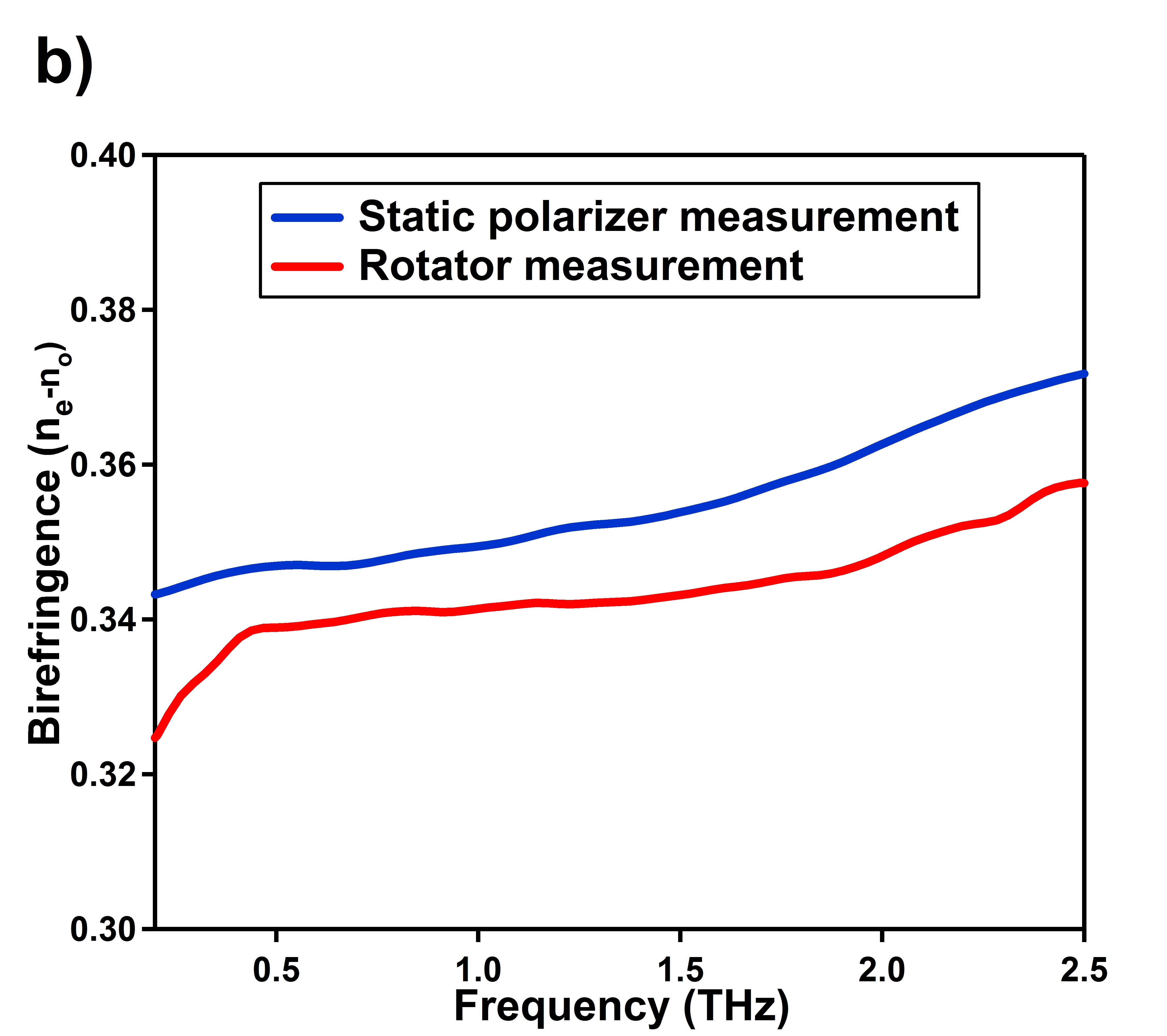} 
\par\end{centering}

\caption{a) Single measurement of the birefringence of sapphire with the electric
field incident on the sample at a $45^{\circ}$ angle to the two principle
axes. b) Calculation of the birefringence ($n_{e}-n_{o}$) for the
ordinary and extraordinary axes for a static polarizer measurement
(requiring 2 measurements plus a reference) and a polarization modulation
measurement (requiring 1 measurement).}

\label{sapphire} 
\end{figure}

As demonstrated above, the accuracy and precision of the technique
improve significantly for smaller rotation angles of the test polarizer.
Fig. \ref{fig:Polarization-angle-characterizat}c shows a measurement
for a polarizer angle of $6{}^{\circ}$. For this small angle, within
the range of 0.1-1.25 THz the measured angle is constant to within
$0.05{}^{\circ}$, and from the signal to noise ratio, we estimate
that changes in the angle as small as $0.02{}^{\circ}$ (350 $\mu$rad)
can be resolved. The main limitation in measuring the accuracy of
the technique is the difficulty in setting the angle of the test polarizer
to the required level of accuracy. At this point we can measure THz
polarization steps much more precisely that we can generate them.
The precision of the system is thus far only limited by the
averaging time. The scans shown in Fig. \ref{fig:Polarization-angle-characterizat}
each account for 20 minutes of measurement, which is reasonable for
many experiments.

To test the technique on a real sample, the birefringent response
of a piece of X-cut sapphire was measured. The sample was placed with
the ordinary and extraordinary axes at $45^{\circ}$ with respect
to the initial vertical ($x$) light polarization. This crystal orientation
projects the electric field of the terahertz waveform equally on the
two crystal axes, producing a phase delay between the two components
and changing the polarization from linear to elliptical. For a purely birefringent
material the $T$ matrix takes the form

\begin{equation}
\left(\begin{array}{cc}
e^{i\varphi_{x}}\cos^{2}\alpha+e^{i\varphi_{y}}\sin^{2}\alpha & \left(e^{i\varphi_{x}}-e^{i\varphi_{y}}\right)\cos\alpha\sin\alpha\\
\left(e^{i\varphi_{x}}-e^{i\varphi_{y}}\right)\cos\alpha\sin\alpha & e^{i\varphi_{x}}\sin^{2}\alpha+e^{i\varphi_{y}}\cos^{2}\alpha
\end{array}\right)\label{eq:birefringence}
\end{equation}

\noindent where $\alpha$ is the angle between the extraordinary axis and laboratory
$x$ axis, and $\varphi_{x}$ and $\varphi_{y}$ are the additional phases
associated with the electric field traveling along the extraordinary
and ordinary axes, respectively. Here, with $\alpha=45^{\circ}$, Eq. (\ref{eq:birefringence})
becomes

\begin{equation}
\frac{1}{2}\left(\begin{array}{cc}
e^{i\varphi_{x}}+e^{i\varphi_{y}} & e^{i\varphi_{x}}-e^{i\varphi_{y}}\\
e^{i\varphi_{x}}-e^{i\varphi_{y}} & e^{i\varphi_{x}}+e^{i\varphi_{y}}
\end{array}\right)\label{eq:birefringence-2}
\end{equation}

For the birefringent response, introducing $\Delta\varphi=\varphi_{x}-\varphi_{y}$,
we find that

\begin{eqnarray}
\frac{S_{y}\left(\omega\right)}{S_{x}\left(\omega\right)} & = & \frac{e^{i\phi_{x}}-e^{i\phi_{y}}}{e^{i\phi_{x}}+e^{i\phi_{y}}}\nonumber\\
\frac{S_{y}\left(\omega\right)}{S_{x}\left(\omega\right)} & = & i\frac{\sin\left(\Delta\varphi/2\right)}{\cos\left(\Delta\varphi/2\right)}
\label{eq:bir-eval}
\end{eqnarray}

meaning the phase difference between the two axes is

\begin{equation}
\Delta\varphi=2\arctan\left(\mathrm{Im}\left[\frac{S_{y}\left(\omega\right)}{S_{x}\left(\omega\right)}\right]\right)
\end{equation}

In Fig. \ref{sapphire}a, we show a time-domain trace of the measured
electric field amplitudes as a function of time, measured directly
on the in-phase and out-of-phase channels of the lock-in amplifier.
It shows quite readily the conversion of the initial linearly polarized
THz pulse into an elliptically polarized state.
From the Fourier transforms of the data and the phase difference
between field components, the frequency dependent difference in the index of refraction
between the ordinary and extraordinary axes can easily be computed
using $\Delta n=c\,\Delta\phi/2\pi df$, where $d$ is the thickness
of the sapphire. In Fig. \ref{sapphire}b we compare the measured birefringence of the sapphire
taken with the conventional technique using static wire grid polarizers and the rotator technique. For the static wire grid polarizer technique, the ordinary axis of the sapphire is oriented along the $x$ axis. A wire grid polarizer oriented at $45^{\circ}$ is placed before the sample, defining the incoming electric field polarization at $45^{\circ}$ to the ordinary and extraordinary axes. An analyzing polarizer is placed after the sample, and two measurements are done at $\pm 45^{\circ}$. A quick mathematical analysis with the tools described above shows that the two measurements then give

\begin{eqnarray}
S_{-45^{\circ}}
& = & 
e^{i\varphi_{x}}-e^{i\varphi_{y}}\nonumber\\
S_{+45^{\circ}}
& = & 
e^{i\varphi_{x}}+e^{i\varphi_{y}}
\label{deg45}
\end{eqnarray}

\noindent thus taking the ratios of these two measurements yields the same result for $\Delta \varphi$ as the rotator measurement.

Fig. \ref{sapphire}b shows that the two measurements produce slightly different results. The difference in the two measurements is ascribed to the imperfect nature of the polarizers.  For the fast rotator experiment, to second order the effects of the finite polarizer extinction ratios drop out. In the static experiment, analysis using the imperfect polarizer (Eq. \ref{finite_polarizer}) changes Eq. (\ref{deg45}) such that $e^{i\phi_{x}}\rightarrow (1+\eta)^{2} e^{i\phi_{x}}$ and $e^{ i\phi_{y}}\rightarrow (1-\eta)^{2} e^{i\phi_{y}}$. The robustness of the polarization modulation technique actually produces a measurement that is less prone to error than that with the static polarizers. This simple measurement demonstrates the power of the technique: a single polarization modulation measurement produces the same information as three measurements with a static polarizer setup, and gives much improved accuracy and precision.

In conclusion, we have demonstrated high precision measurement of
polarization states using time domain terahertz spectroscopy. We can
resolve angular rotations with an accuracy of $\sim0.05^{\circ}$
and a precision of $\sim0.02^{\circ}$. Additionally, a number of
practical mathematical results useful in the analysis of phase sensitive
polarization measurements have been presented. We believe that this
technique will have a wide applicability to a number of materials
systems at the forefront of condensed matter physics, such as high-T$_c$
superconductors, quantum magnets, and topological insulators.

\section*{Acknowledgements}

The authors would like to thank J. Orndorff of the Instrument Development Group in the Johns Hopkins Department of Physics and Astronomy for the development and construction of the fast rotator. The authors would also like to thank M. Neshat, A. Marklez, J. Cerne, and G. Jenkins for helpful discussions.  This work was made possible by support from the Gordon and Betty Moore Foundation and DARPA YFA N66001-10-1-4017


\begin{thebibliography}{10}

\bibitem{Tonouchi2007} M. Tonouchi, ``Cutting-edge terahertz technology," Nature Photonics {\bf 1}, 97 (2007).

\bibitem{Nuss1998} M. Nuss and J. Orenstein, ``Terahertz time-domain spectroscopy" in {\em Millimeter and Submillimeter Wave Spectroscopy of Solids}, George Gr\"{u}ner, ed., (Springer Berlin / Heidelberg, 1998), Topics in Applied Physics {\bf 74}, 7--50.

\bibitem{DOEreport} M. S. Sherwin, C. A. Schmuttenmaer, and P. H. Bucksbaum, eds. ``Opportunities in THz science," DOE-NSF-NIH Workshop, Feb. 12-14, 2004, $http://science.energy.gov/\sim/media/bes/pdf/reports/files/thz\_rpt.pdf$.

\bibitem{Kaindl2003} R. A. Kaindl, M. A. Carnahan, D. H\"{a}gele, R. L\"{o}venich, and D. S. Chemla, ``Ultrafast terahertz probes of transient conducting and insulating phases in an electron-hole gas," Nature {\bf 423}, 734 (2003).

\bibitem{Heyman1998}J. N. Heyman, R. Kersting, and K. Unterrainer, ``Time-domain measurement of intersubband oscillations in a quantum well," \apl {\bf 72}, 644--646 (1998).

\bibitem{coatings} A.J. Gatesman, J. Waldman, M. Ji, C. Musante, and S. Yagvesson, ``An anti-reflection coating for silicon optics at terahertz frequencies," Microwave and Guided Wave Letters, IEEE {\bf 10}, 264--266 (2000).

\bibitem{Bilbro2011} L. S. Bilbro, R. Vald\'{e}s Aguilar, G. Logvenov, O. Pelleg, I. Bozovic, and N. P. Armitage, ``Temporal correlations of superconductivity above the transition temperature in $\mathrm{La_{2-x}Sr_xCuO_4}$ probed by terahertz spectroscopy," Nature Physics {\bf 7}, 298--302 (2011).

\bibitem{Corson1999} J. Corson, R. Mallozzi, J. Orenstein, J. N. Eckstein, and I. Bozovic, ``Vanishing of phase coherence in underdoped $\mathrm{Bi_2Sr_2CaCu_2O_{8+\delta}}$," Nature {\bf 398}, 221 (1999).

\bibitem{THzBiohazard} S. Wang, B. Ferguson, D. Abbott, and X.-C. Zhang, ``T-ray imaging and tomography," Journal of Biological Physics {\bf 29}, 247--256 (2003).

\bibitem{THzproteins} M. Nagel, P. H. Bolivar, M. Brucherseifer, H. Kurz, A. Bosserhoff, and R. B\"{u}ttner, ``Integrated planar terahertz resonators for femtomolar sensitivity label-free detection of dna hybridization," Appl. Opt. {\bf 41}, 2074--2078 (2002).

\bibitem{THzimaging0}B. B. Hu and M. C. Nuss, ``Imaging with terahertz waves," Opt. Lett. {\bf 20}, 1716--1718 (1995).

\bibitem{THzimaging1} X-C Zhang, ``Terahertz wave imaging: horizons and hurdles," Physics in Medicine and Biology {\bf 47}, 3667 (2002).

\bibitem{THzimaging3} J. L. Johnson, T. D. Dorney, and D. M. Mittleman, `` Enhanced depth resolution in terahertz imaging using phase-shift interferometry,"  \apl {\bf 78}, 835--837 (2001).

\bibitem{Castro-Camus} E.~Castro-Camus, ``Polarization-resolved terahertz time-domain spectroscopy," Journal of Infrared, Millimeter and Terahertz Waves, 1--13 (2011).

\bibitem{Shimano2011} R.~Shimano, Y.~Ikebe, K.~S. Takahashi, M.~Kawasaki, N.~Nagaosa, and Y.~Tokura, ``Terahertz Faraday rotation induced by an anomalous Hall effect in the itinerant ferromagnet $\mathrm{SrRuO_{3}}$," Europhysics Letters {\bf 95}, 17002 (2011).

\bibitem{Kapitulnik1994} A.~Kapitulnik, J.~S. Dodge, and M.~M. Fejer, ``High-resolution magneto-optic measurements with a sagnac interferometer," Journal of Applied Physics {\bf 75}, 6872--6877 (1994).

\bibitem{Xia2006}J. Xia, Y. Maeno, P. T. Beyersdorf, M. M. Fejer, and A. Kapitulnik, ``High resolution polar kerr effect measurements of ${\mathrm{Sr}}_{2}{\mathrm{RuO}}_{4}$: Evidence for broken time-reversal symmetry in the superconducting state," Phys. Rev. Lett. {\bf 97}, 167002 (2006).

\bibitem{Castro-Camus2005} E. Castro-Camus, J. Lloyd-Hughes, M. B. Johnston, M. D. Fraser, H. H. Tan, and C. Jagadish, ``Polarization-sensitive terahertz detection by multicontact photoconductive receivers," \apl {\bf 86}, 254102 (2005).

\bibitem{Makabe2007} H. Makabe, Y. Hirota, M. Tani, and M. Hangyo, ``Polarization state measurement of terahertz electromagnetic radiation by three-contact photoconductive antenna," Opt. Express {\bf 15}, 11650--11657 (2007).

\bibitem{Neshat12a} M.~{Neshat} and N.~P. {Armitage}, ``Improved measurement of polarization state in terahertz polarization spectroscopy," In press, Optics Letters (2012).

\bibitem{Qi2008} X.-L. Qi, T. L. Hughes, and S.-C. Zhang, ``Topological field theory of time-reversal invariant insulators" Phys. Rev. B {\bf 78}, 195424 (2008).

\bibitem{Nandkishore2012} R. Nandkishore, L. S. Levitov, and A. V. Chubukov, ``Chiral superconductivity from repulsive interactions in doped graphene," Nature Physics {\bf 8}, 158--163 (2012).

\bibitem{Tse11a} W.-K. Tse and A. H. MacDonald, ``Magneto-optical Faraday and Kerr effects in topological insulator films and in other layered quantized Hall systems," Phys. Rev. B {\bf 84}, 205327 (2011).

\bibitem{Maciejko10a} J. Maciejko, X.-L. Qi, H. D. Drew, and S.-C. Zhang, ``Topological quantization in units of the fine structure constant," Phys. Rev. Lett. {\bf 105}, 166803 (2010).

\bibitem{Shan2009} J. Shan, J. I. Dadap, and T. F. Heinz, ``Circularly polarized light in the single-cycle limit: The nature of highly polychromatic radiation of defined polarization," Opt. Express {\bf 17}, 7431--7439 (2009).

\bibitem{Bernhard2004} C. Bernhard, J. Huml{\i}cek, and B. Keimer, ``Far-infrared ellipsometry using a synchrotron light source-the dielectric response of the cuprate high $\mathrm{T_c}$ superconductors," Thin solid films {\bf 455}, 143--149 (2004).

\bibitem{Grayson2002} M. Grayson, L. B. Rigal, D. C. Schmadel, H. D. Drew, and P.-J. Kung, ``Spectral measurement of the Hall angle response in normal state cuprate superconductors," Phys. Rev. Lett. {\bf 89}, 037003 (2002).

\bibitem{Jenkins2010} G. S. Jenkins, D. C. Schmadel, and H. D. Drew, ``Simultaneous measurement of circular dichroism and Faraday rotation at terahertz frequencies utilizing electric field sensitive detection via polarization modulation," Review of Scientific Instruments {\bf 81}, 083903 (2010).

\bibitem{RVA2012} R. Vald\'{e}s Aguilar, A. V. Stier, W. Liu, L. S. Bilbro, D. K. George, N. Bansal, L. Wu, J. Cerne, A. G. Markelz, S. Oh, N. P. Armitager, ``Thz response and colossal kerr rotation from the surface states of the topological insulator Bi$_2$Se$_3$," Physical Review Letters {\bf 108}, 087403 (2012).

\bibitem{George12} D. K. George, A. V. Stier, C. T. Ellis, B. D. McCombe, J. {\v C}erne, and A. G. Markelz, ``Terahertz Magneto Optical Polarization Modulation Spectroscopy," arXiv:1201.2701v1 [cond-mat.mtrl-sci] (2012).

\bibitem{Huang2010} C. Huang, S. Zhao, H. Chen, and Z. Liao, ``Attenuation characterization of multiple combinations of imperfect polarizers," J. Opt. Soc. Am. A {\bf 27}, 1060--1068 (2010).

\bibitem{Ren0} L. Ren, C. L. Pint, T. Arikawa, K. Takeya, I. Kawayama, M. Tonouchi, R. H. Hauge, and J. Kono, ``Broadband terahertz polarizers with ideal performance based on aligned carbon nanotube stacks," Nano Letters, {\bf 12}, 787 (2012).

\bibitem{fujiwara}H. Fujiwara, ``Spectroscopic Ellipsometry: Principles and Applications," (John Wiley \& Sons, Ltd., 2006).

\bibitem{Born1997} M. Born and E. Wolf, ``Principles of Optics," (Cambridge University Press, 1997).

\bibitem{Masson2006} J.-B. Masson and G. Gallot, ``Terahertz achromatic quarter-wave plate," Opt. Lett. {\bf 31}, 265--267 (2006).

\end{thebibliography}
\end{document}